\documentclass[review]{elsarticle}

\usepackage{lineno,amsfonts,longtable}
\modulolinenumbers[5]

\journal{Computational in Biology and Chemistry}

\begin{document}

\begin{frontmatter}

\title{Deep Belief Network based representation learning for lncRNA-disease association prediction}


\author[mymainaddress,mysecondaryaddress]{Manu Madhavan \corref{mycorrespondingauthor}}
\cortext[mycorrespondingauthor]{Corresponding author}
\ead{kkmanumadhavan@gmail.com}

\author[mymainaddress]{Gopakumar G}

\ead{gopakumarg@nitc.ac.in}

\address[mymainaddress]{Department of Computer Science and Engineering, National Institute of Technology Calicut, 673601, India}
\address[mysecondaryaddress]{Department of Computer Science and Engineering, Amrita School of Engineering, Coimbatore, Amrita Vishwa Vidyapeetham, 641112, India.}

\begin{abstract}
\textbf{Background:} The expanding research in the field of long non-coding RNAs (lncRNAs) showed abnormal expression of lncRNAs in many complex diseases. Accurately identifying lncRNA-disease association is essential in understanding lncRNA functionality and disease mechanism. There are many  machine learning techniques involved in the prediction of lncRNA-disease association which use different biological interaction networks and associated features. Feature learning from the network structured data is one of the limiting factors of machine learning-based methods. Graph neural network based techniques solve this limitation by unsupervised feature learning. Deep belief networks (DBN) are recently used in biological network analysis to learn the latent representations of network features.  \\
\textbf{Method: } In this paper, we propose a DBN based lncRNA-disease association prediction model (DBNLDA) from lncRNA, disease and miRNA interactions. The architecture contains three major modules- network construction, DBN based feature learning and neural network-based prediction. First, we constructed three heterogeneous networks such as lncRNA-miRNA similarity (LMS), disease-miRNA similarity (DMS) and lncRNA-disease association (LDA) network. From the node embedding matrices of similarity networks, lncRNA-disease representations were learned separately by two DBN based subnetworks. The joint representation of lncRNA-disease was learned by a third DBN from outputs of the two subnetworks mentioned. This joint feature representation was used to predict the association score by an ANN classifier. \\
\textbf{Results: }The proposed method obtained AUC of 0.96 and AUPR of 0.967 when tested against standard dataset used by the state-of-the-art methods. Analysis on breast, lung and stomach cancer cases also affirmed the effectiveness of DBNLDA in predicting significant lncRNA-disease associations.\\
\textbf{Availability:} https://github.com/manumad/DBNLDA

\end{abstract}


\end{frontmatter}


\section{Introduction} \label{sec:intro}
Long non-coding RNAs (lncRNA) are RNA transcripts with more than 200 nucleotides in length and lack of protein-coding potential  \cite{goff2015linking}. The studies revealed that lncRNAs were involved as a regulator in many biological processes such as ageing, cell differentiation, epigenetic mechanisms and protein synthesis  \cite{lee2012epigenetic}. The aberrant expression of lncRNAs is also associated with many complex diseases like cancers \cite{huarte2015emerging}, alzheimer\rq s disease \cite{luo2016long}, and heart failure \cite{wang2019construction}.   
Therefore, identifying the role of lncRNAs in diseases will help to improve the understanding of the disease mechanisms and derive new insights on drug therapeutics \cite{bhartiya2012conceptual}.

An array of computational models to predict the lncRNA-disease association are available in the literature with varying performance. All these works utilise varieties of lncRNA features from known lncRNA-disease associations and their interactions with other molecules like micorRNAs (miRNA), proteins and messengerRNAs (mRNA). There are three major categories of works in lncRNA-disease association. The first category of works utilise the knowledge of lncRNA functional similarity under the assumption that functionally similar lncRNAs associate with similar diseases. Based on this assumption, an lncRNA-disease association network was constructed. Then, algorithms from machine learning and social network analysis were used to make the lncRNA-disease prediction. For example, methods such as RWRlncD \cite{sun2014inferring}, IRWRLDA \cite{chen2016irwrlda}, BRWLDA \cite{yu2017brwlda} used various random walk algorithm and KATZLDA \cite{chen2015katzlda} applied Katz page ranking algorithm for  analysing the similarity network. All these methods relied upon the network structure features and results were biased towards nodes having high degree and centrality. 

The advanced research on lncRNA mechanism revealed that the regulation of lncRNA is largely determined by co-expressed miRNAs \cite{paraskevopoulou2016analyzing}. The second category of works used expression levels of lncRNAs, genes and miRNAs in various diseases. The earlier methods \cite{rflda} in this category used experimentally validated disease associated genes/miRNAs and lncRNA co-expression data. These methods were not useful with lncRNAs which have no experimentally validated gene/miRNA interaction. Recent works under this category used mathematical models such as matrix completion \cite{lu2018simclda}, matrix factorization \cite{fu2018mflda} and Graph-based algorithms (TPGLDA \cite{ding2018tpglda}, DisLncPri \cite{wang2017improved}) to predict lncRNA-disease association. 

The third category of works constructed a heterogeneous interaction network based on lncRNA, miRNA and mRNA functional similarity and associations. Machine learning techniques such as Random Walk \cite{rflda}, Support Vector Machine (SVM)  \cite{lan2017ldap}, and Laplacian regularised least square method \cite{lrlslda} have found use in the analysis of these networks. Major challenge in the above methods was the effective representation of lncRNA-disease features. Introduction of deep learning models eliminated the need for feature extraction by enabling unsupervised representation learning. Among them,  CNNLDA \cite{xuan2019dual} used convolutional neural networks and GCNLDA \cite{gcnlda} used graph convolutional neural networks \cite{kipf2016semi} to learn global representations of lncRNA, miRNA and disease nodes. A recent work titled GAMCLDA \cite{wu2020inferring} used graph autoencoder and matrix completion to predict lncRNA-disease association. 

Deep learning algorithms such as Deep Belief Networks (DBN) were used recently in computer vision and text mining to learn the latent representation of the data \cite{hinton2009deep}. A DBN model has stacked layers of Restricted Boltzman Machines (RBM), which contains visible layers and hidden layers to compute probability distribution as latent representation. DBN based models were successfully used in Bioinformatics to predict drug-target \cite{wang2013predicting}, multiple types of miRNA-disease association \cite{chen2015rbmmmda, dgmdl} and cancer sub-type prediction \cite{liang2014integrative}, but not applied in the field of lncRNA-disease association prediction.

\begin{figure}
\includegraphics[scale = 0.2]{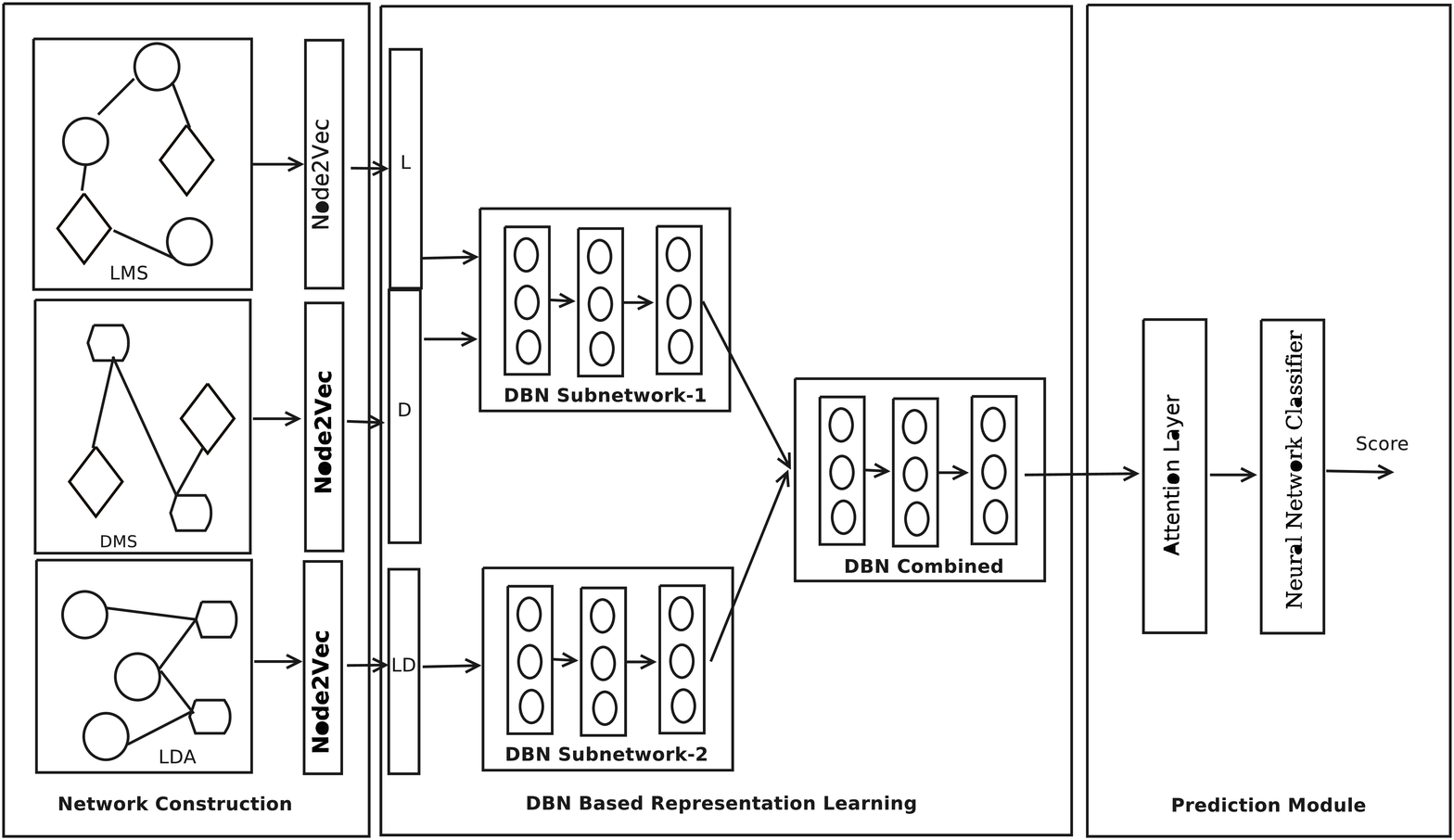}
\caption{Architecture of DBNLDA. LMS:LncRNA-miRNA similarity; DMS: Disease-miRNA similarity; LDA: LncRNA-Disease Association;  L:embedding of lncRNAs from LMS; D: embedding of diseases from DMS; LD: embedding of lncRNAs and diseases from LDA; DBN: Deep Belief Network.} 
\label{fig:arch}
\end{figure}

In this study, we propose a DBN based representation learning model (called DBNLDA) for lncRNA-disease association prediction. DBNLDA make use of heterogeneous information on functional similarity, co-expression and interactions between lncRNAs and diseases for making the prediction. The architecture of DBNLDA contains 3 modules namely-(i)network construction, (ii) DBN based feature learning and (iii) association prediction. Summary of the proposed architecture is depicted in Figure~\ref{fig:arch}. Instead of using a single heterogeneous network of lncRNAs, miRNAs and diseases as in previous works \cite{rflda, gcnlda}, DBNLDA constructs three functional similarity networks such as lncRNA-miRNA (LMS), disease-miRNA (DMS) and lncRNA-Disease association (LDA). Then, for each pair of lncRNA-disease, {\em{DBN  subnetwork-1}} learns lncRNA and disease representations from their functional associations with miRNAs. Similarly, {\em{DBN subnetwork-2}} learns lncRNA-disease representations from LDA network. The representations learned by subnetworks are combined for learning higher-level representation using the third DBN ({\em{DBN-combined}}). In order to reduce the obstructive impact of features learned from sparse network, the features from {\em{DBN-combined}} are recomputed using an attention layer. In the final stage, a neural network-based classifier is used to predict the lncRNA-disease association based on the joint feature representation.  A 5-fold cross-validation and case studies on cancer dataset showed that DBNLDA significantly improves the performance and potential lncRNA-disease association prediction.

The remainder of the paper is organized as follows. Section~\ref{method} gives a detailed discussion of DBNLDA architecture. Section~\ref{results} discusses the results of the proposed method and Section \ref{conclusion} concludes the paper.

\section{Materials and Methods} \label{method}
\subsection{The Dataset}
Datasets used in the previous reports \cite{rflda, gcnlda, fu2018mflda} were used for lncRNA-Disease associations, lncRNA-miRNA interaction, lncRNA function similarity and disease semantic similarity. LncRNA-disease associations were downloaded from two reference databases: LncRNADisease \cite{chen2012lncrnadisease} and lnc2cancer \cite{ning2016lnc2cancer}. MiRNA-lncRNA interactions and miRNA-disease interactions were downloaded from miRNet \cite{fan2016mirnet} and Starbase \cite{li2014starbase} databases respectively. All these downloaded associations and similarities were then compiled for 240 lncRNAs, 412 diseases, 495 miRNAs and 2697 known lncRNA- disease interactions. These known interaction pairs constituted the positive example for training the model. All other pairs between lncRNAs and diseases, which were not listed in reference databases, were considered to be negative interactions. We randomly selected 2967 samples from negative interactions to construct a balanced dataset. Summary of the dataset is given in Table \ref{Tab:dataset}.

\begin{table}[h]
\centering
\caption{Summary of dataset}\label{Tab:dataset}
\begin{tabular}{ c c } 
\hline
Particular& Count\\
\hline
LncRNAs &290\\
Diseases &412\\
MiRNAs &495\\
Positive pairs    &2697\\
Negative pairs    &2697\\
\hline
\end{tabular}
\end{table}

\subsection{Construction of LMS, DMS and LDA networks}
The first step in the DBNLDA architecture was the construction of three similarity networks such as LMS, DMS and LDA as defined in Section \ref{sec:intro}. Let $nl, nm,$ and $nd$ be the number of lncRNAs, miRNAs and diseases in the dataset. LMS network (with $nl$ lncRNAs and $nm$ miRNAs) was constructed using lncRNA-lncRNA similarity and lncRNA-miRNA interactions. LncRNA functional similarities computed by Chen's method \cite{chen2015constructing} was adapted here, and an edge was added if the similarity score was greater than 0. For the list of lncRNAs, the known miRNA targets downloaded from the miRNet were used as lncRNA-miRNA edges. 

Similar to LMS, DMS network was constructed (with $nd$ diseases and $nm$ miRNAs) using disease-disease similarities and known disease-miRNA associations. LncRNA-Disease association (LDA) network contains $nl$ lncRNAs and $nd$ diseases, where an un-directed edge was used to represent the known association between lncRNA and disease.

\subsection{Computing node embedding features from networks}
Node2vec \cite{grover2016node2vec} algorithm was used to get node embedding from each network. Node2vec output a node representation based on the neighbourhood information of the node.  First, LMS network was passed through node2vec embedding layer, and embedding matrix for lncRNA was obtained as $L_{lms} \in \mathbb{R}^{nl\times e}$, where $e$ is the embedding dimension.  The embedding matrix for diseases from DMS was obtained as $D_{dms} \in \mathbb{R}^{nd\times e}$. Then, for each lncRNA-disease pair $(l,d)$ in the dataset (both positive and negative examples), $L_{lms}[l]$ and $D_{dms}[d]$ were concatenated to form vector $LD_{lmd}[ld]\in \mathbb{R}^{2e}$. This resulted in a feature matrix $LD_{lmd}\in \mathbb{R}^{n \times 2e}$, where $n$ is the total number of examples in the dataset. Finally, the embedding matrices of lncRNAs and diseases, represented as $L_{lda}\in \mathbb{R}^{nl\times e}$, $D_{lda}\in \mathbb{R}^{nd \times e}$ were obtained from LDA network. Then as in the case of $LD_{lmd}$, for each pairs of lncRNA-disease, vectors from $L_{lda}$ and $D_{lda}$ were concatenated to form feature matrix $LD_{lda}\in \mathbb{R}^{n\times 2e}$. 

\subsection{DBN based feature learning}
Following the Node2vec embedding layer, DBNLDA implements Deep Belief Network based feature learning. In this work, DBN was used to learn latent representation of lncRNA and disease nodes and encode them to a new dimension $h$, where $h\geq e$. The architecture consisted of two DBN subnetworks- one to learn lncRNA-disease representation from functional similarity networks (LMS and DMS) and other from LDA network. The {\em{DBN subnetwork-1}} received embedded feature matrix $LD_{lmd}$ as input, and produce $LD_{db1}\in \mathbb{R}^{n\times h}$ ($h$, the number of hidden units in DBN) as output. Similarly, the {\em{DBN subnetwork-2}} received $LD_{lda}$ as input and produce $LD_{db2} \in \mathbb{R}^{n\times h}$ as output. Then both $LD_{db1}$ and $LD_{db2}$ were concatenated to form combined feature representation, $LD\in \mathbb{R}^{n\times 2h}$.  The network {\em{DBN-combined}} accepted $LD$ as input and learn the feature representation as $LD_{db} \in \mathbb{R} ^n\times h'$, where $h'$ is the number of hidden units in the {\em{DBN-combined}} and $h' \geq h$.  
 
\subsection{Feature attention layer}
Since all the three networks used in the DBNLDA architecture were sparse in nature, there was a chance of losing feature importance in the latent representation learned by DBN. The attention mechanism in deep learning was used to solve this issue by recomputing the feature values from all available information so that their contributions could be different and unique. In this work, attention mechanism similar to the one used in GCNLDA  \cite{gcnlda} was applied to $LD_{db}$ features. 

Let $LD_{db_{i}}= [ld_{i}^{1}, ld_{i}^{2}, ..., ld_{i}^{h'}]$ represents the feature vector of $i^{th}$ entry in the dataset learned by {\em{DBN-combined}}, where $ld_{i}^k \in \mathbb{R}, \forall k=1,2,...h'$. The attention score for each element in $LD_{db_i}$ was calculated by introducing attention weight parameters $H^{att} \in \mathbb{R}^{h'\times h'}$, $W^{att}\in \mathbb{R}^{h'\times h'}$ and  bias $b^{att} \in \mathbb{R}^{h'}$ as in Equation \ref{equ1}.

\begin{equation} \label{equ1}
 \alpha_{i}^{att} = Softmax(H^{att} \cdot tanh(W^{att} LD_{i} +b^{att}))
\end{equation}

Next, the attention enhanced feature values were recomputed as in 
Equation \ref{equ2}.

\begin{equation} \label{equ2}
 LD_{i}^{att} = \alpha_{i}^{att} \otimes LD_{db_i}
\end{equation}

where $\otimes$ represents pairwise multiplication. Finally, the matrix $LD^{att} = [LD_{i}^{att}]$ was used as input for association prediction.

\subsection{Prediction Layer}
The final module of DBNLDA architecture is a neural network classifier to predict the association between lncRNA and disease. The classifier network followed a 1-2-1 structure, where the input layer has $h'$ nodes to receive input from $LD^{att}$ and output layer has one node to predict the association score. The number of nodes in hidden layers kept as hyperparameter. Activation function used in all layers except the output was Relu. In order to reduce the overfitting, a dropout of 0.02 probability was added between the hidden layers. Output layer used a sigmoid function to compute the association score. The network used ADAM optimizer with learning rate 0.01 and binary cross-entropy loss function. 

\subsection{Hyperparameters}
\begin{figure}
\includegraphics[scale = 0.85]{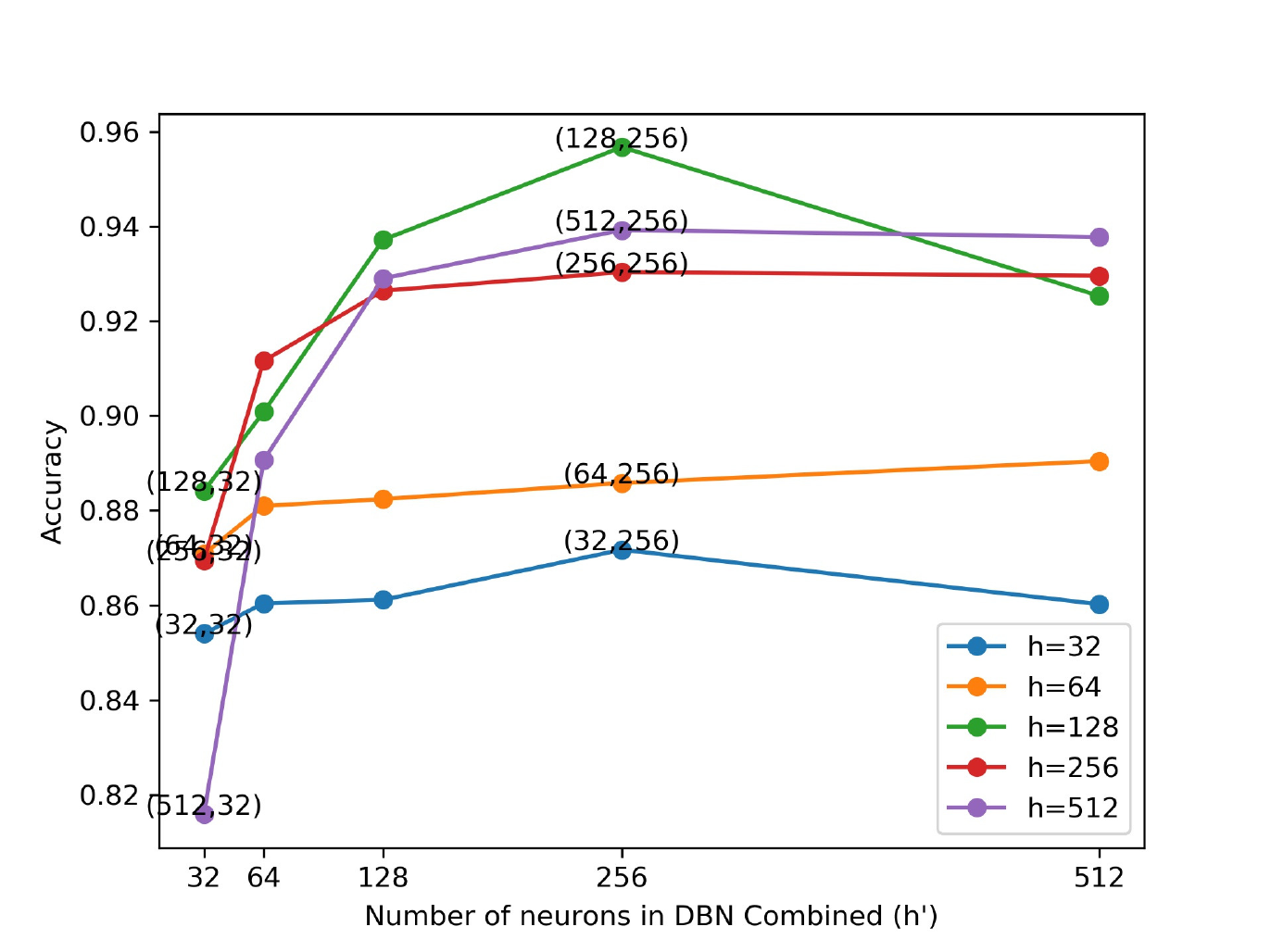}
\caption{Comparison of accuracy of DBNLDA model on different number of nodes in DBN subnetworks ($h$) and DBN combined ($h'$). Points are annotated with ($h$, $h'$) values} 
\label{fig:hyper}
\end{figure}

Various hyperparameters determine the performance of DBNLDA in different modules. The values of these hyperparameters were set empirically following performance evaluation. The dimension of node2vec embedding, $e$ is set to be 64 for all networks with keeping other parameters to default values in \cite{grover2016node2vec}. Following the implementation of \cite{dgmdl}, the DBN subnetworks in this work uses three stacked layers of RBM, with $h=128$ nodes in all layers. The combined DBN has dimension $h'=256$. It is found that very low values $h$ and $h'$ degrade the performance and high values have no effect on the performance (see Figure ~\ref{fig:hyper} for details). 
For the classification module, the number of neurons in both hidden layers set as 128. The classifier iterate over 30 epochs, since the model performance became stable after $30^{th}$ epoch.

A detailed description DBNLDA steps and implementation details are available in SupplementaryFile-S1.pdf.

\section{Results and Discussion} \label{results}

\subsection{Performance evaluation metrics}
We used 5-fold cross-validation to measure and compare the performance of the model. The dataset consist of 5394 lncRNA-disease pairs (2697 positive associations and 2697 negative associations). The area under the receiver operating characteristic (AUC-ROC) curve was used to compare the global performance of the prediction model. The area under the precision-recall curve (AUPR) and average accuracy were also used to measure the prediction performance.   

\subsection{Overall performance}
\begin{figure}
\centering
\includegraphics[scale = 0.8]{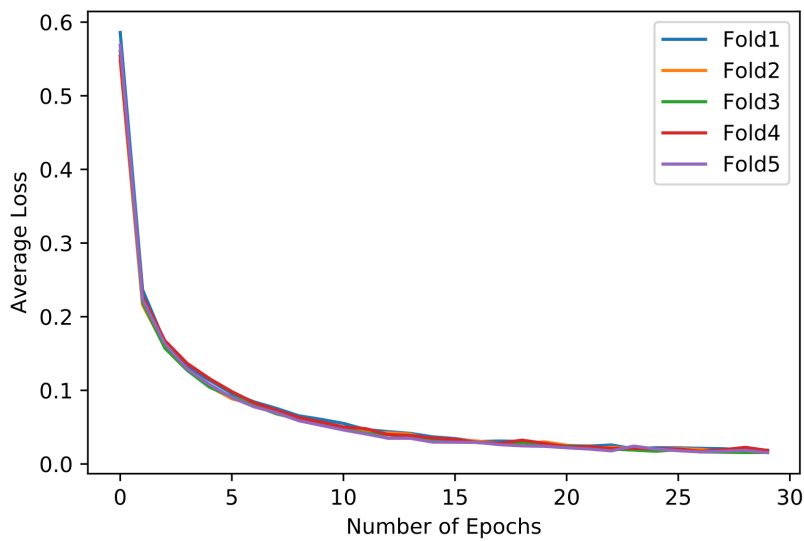}
\caption{Learning Curve for different folds of 5-fold cross-validation} 
\label{fig:learning}
\end{figure}

\begin{figure}
\includegraphics[scale=0.8]{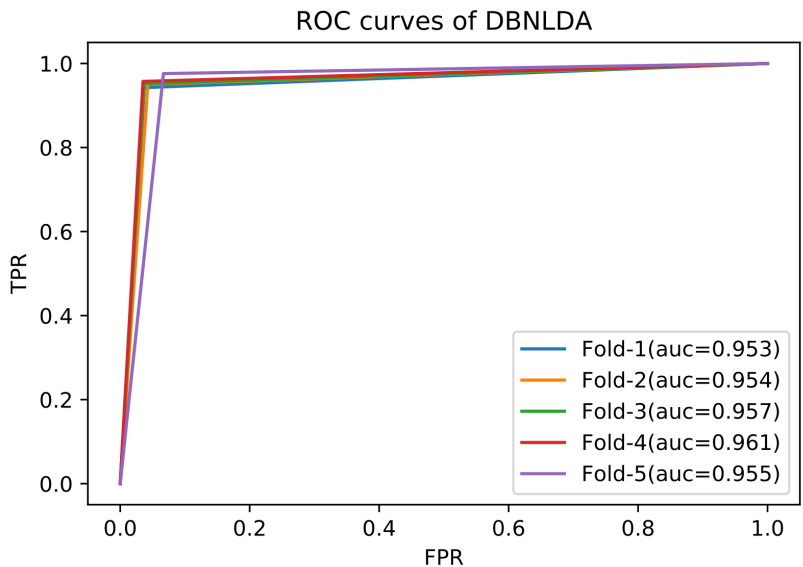}
\caption{ROC curve for different folds of 5-fold cross-validation} \label{fig:roc}
\end{figure}

The DBNLDA was trained over 30 epochs and the learning curve (refer Figure~\ref{fig:learning}) showed consistent characteristics in all folds of cross validation. The model gave average AUC of 0.96 over cross validations and ROC curve is shown in Figure~\ref{fig:roc}. The model reported an accuracy of 0.957 and AUPR value of 0.968. 

In order to analyse how DBN based features improve the performance of the prediction model, we have repeated the experiments with different levels of feature combinations. It is clear from the Table~\ref{Tab:feature} that introduction of DBN based learning significantly improved the accuracy of the prediction model.

\begin{table}[h]
\centering
\caption{Accuracy of the model based on feature combinations}\label{Tab:feature}
\begin{tabular}{c l l}
\hline
Experiment & Feature & Accuracy \\
\hline
Exp 1 &only Node2vec featues & 0.817\\
Exp 2 &Node2vec, {\em{DBN subnetwork-1}} and &0.896\\
&{\em{DBN subnetwork-2}} &\\
Exp 3 &Node2vec, {\em{DBN subnetwork-1}}, & 0.956\\
& {\em{DBN subnetwork-2}} and {\em{DBN-combined}}& \\
\hline
\end{tabular}
\end{table}

\subsection{Comparison with other methods}
We compared the performance of DBNLDA with other state-of-the-art methods such as RFLDA \cite{rflda}, GCNLDA \cite{gcnlda}, SIMCLDA\cite {lu2018simclda}, Ping\rq s Method \cite{ping2018novel}, MFLDA  \cite{fu2018mflda}, LDAP \cite{lan2017ldap}, GAMCLDA \cite{wu2020inferring} and CNNLDA \cite{xuan2019dual}, based on area under ROC curve (AUC) and Area under Precision-Recall curve (AUPR). The above methods used knowledge from heterogeneous information from different data sources  to predict lncRNA-disease association and not consider network structure features. AUC and AUPR values of all LDA prediction models were given in Table~\ref{Tab:comp}. The AUC and AUPR values of all methods except DBNLDA were taken from \cite{rflda} and \cite{wu2020inferring}. It was evident from the table that DBNLDA reported second best AUC (0.96) which is closer (1.6\% less) to the highest AUC value  reported by RFLDA. Moreover, DBNLDA outperforms all other methods in terms of AUPR values, where DBNLDA have 0.968 which is 18.9\% better than the second highest value (RFLDA). These results show that DBNLDA predicts lncRNA-disease associations effectively.

\begin{table}[h]
\centering
\caption{Comparison of performance with state-of-the-art methods}
\label{Tab:comp}
\begin{tabular}{l l l} 
\hline 
Method & AUC & AUPR\\
\hline
MFLDA \cite{fu2018mflda}  &0.626 &0.066\\
SIMCLDA \cite{lu2018simclda} &0.746 &0.095\\
LDAP \cite{lan2017ldap}  &0.863 &0.166\\
Ping\rq s Method \cite{ping2018novel} &0.871 &0.219\\
GAMCLDA \cite{wu2020inferring} & 0.907 &0.037\\
CNNLDA \cite{xuan2019dual}  &0.952 &0.251\\
GCNLDA \cite{gcnlda} & 0.959 &0.223\\
RFLDA \cite{rflda} &0.976 &0.779\\
\textbf{DBNLDA} &\textbf{0.960} &\textbf{0.968}\\
\hline
\end{tabular}
\end{table}

\subsection{Case studies}
\begin{table}[h!]
\caption{Top 15 DBNLDA predicted lncRNAs associated with breast cancer}
\label{Tab:cancer1}
\begin{tabular}{ l l l}
\hline
lncRNA &Rank &~Evidence\\
\hline
GAS5 &1 & Lnc2Cancer \\
DLEU2 &2 &LncRNADisease \\ 
HCP5 &3 &Literature (PMID: 31864836) \cite{wu2019downregulation} \\
HOTAIR &4 &LncRNADisease, Lnc2Cancer \\
MEG3 &5 & LncRNADisease, Lnc2Cancer\\
HULC &6 & Lnc2Cancer\\
BCYRN1 &7 &LncRNADisease \\
HOTTIP &8 & Lnc2Cancer\\
UCA1 &9 & LncRNADisease, Lnc2Cancer\\
CDKN2B-AS1 &10 & LncRNADisease\\
NEAT1 &11 & LncRNADisease, Lnc2Cancer\\
TUG1 &12 & LncRNADisease, Lnc2Cancer\\
AFAP1-AS1 &13 & Lnc2Cancer\\
MIR100HG &14 & Literature (PMID:30042378) \cite{wang2018lncrna}\\
TINCR &15 & Lnc2Cancer\\
\hline
\end{tabular}
\end{table}

\begin{table*}[h!]
\caption{Top 15 DBNLDA predicted lncRNAs associated with lung cancer}
\label{Tab:cancer2}
\begin{tabular}{l l l}
\hline
lncRNA &Rank &~Evidence\\
\hline
TUG1	&1 & Literature (PMID:31532756) \cite{guo2019long}\\
PVT1	&2 & Lnc2Cancer\\
AFAP1-AS1 &3 & LncRNADiease, Lnc2Cancer\\
XIST	&4 & Literature (PMID: 28448993) \cite{wang2017long}\\
CCAT2	&5 & LncRNADisease\\
MALAT1	&6 & LncRNADiease, Lnc2Cancer\\
HOTTIP	&7 & LncRNADiease, Lnc2Cancer\\
SOX2-OT	&8 & LncRNADiease\\
HULC	&9 & Literature (PMID:30575912) \cite{liu2018lncrna}\\
MIR155HG &10 & Literature (PMID:32129458) \cite{song2020lncrna}\\
CDKN2B-AS1 &11 & Literature (PMID:29541247) \cite{du2018low}\\
BANCR	&12 & LncRNADiease, Lnc2Cancer\\
BCYRN1	&13 & LncRNADiease\\
UCA1	&14 & LncRNADiease, Lnc2Cancer\\
H19	&15 & LncRNADiease, Lnc2Cancer\\
\hline
\end{tabular}
\end{table*}

\begin{table*}[h!]
\caption{Top 15 DBNLDA predicted lncRNAs associated with stomach cancer.}
\label{Tab:cancer3}
\begin{tabular}{l l l}
\hline
lncRNA &Rank &~Evidence\\
\hline
MIR17HG &1 &Literature (PMID:26837962) \cite{bahari2015mir} \\ 
BCYRN1 &2  &Literature (PMID:29435146) \cite{ren2018upregulation}\\
BANCR &3  &LncRNADisease\\
HCP5 &4 &*\\
AFAP1-AS1 &5 &LncRNADisease\\   
HNF1A-AS1 &6 &LncRNADisease, Lnc2Cancer\\
NALT1 &7   &Literature (PMID:31802831) \cite{piao2019long} \\
DANCR &8 &Lnc2Cancer\\
MIR99AHG &9 &*\\
GAS5 &10 &LncRNADisease, Lnc2Cancer\\
HULC &11 &LncRNADisease, Lnc2Cancer\\
HCG4 &12 &*\\
XIST &13 &LncRNADisease, Lnc2Cancer\\
HOTTIP &14 &LncRNADisease, Lnc2Cancer\\
UCA1 &15 &LncRNADisease, Lnc2Cancer\\
\hline
\end{tabular}
\end{table*}

To further investigate the ability of DBNLDA in predicting significant lncRNA-disease associations, case studies on breast cancer, lung cancer and stomach cancer were conducted. For this study, first trained the DBNLDA model on a dataset containing all lncRNA-disease associations except the validated association between lncRNAs and the disease of interest (breast/lung/stomach cancer). Then the association score for all lncRNAs to the particular disease was calculated using the trained model and analysed the top 15 candidate lncRNAs for each disease. Tables~\ref{Tab:cancer1},\ref{Tab:cancer2}, and \ref{Tab:cancer3} show the top 15 candidate lncRNAs in breast/lung/stomach cancer,  predicted by DBNLDA. The evidence column shows the reference to the associations either from reference databases or literature.  

It was found that 13 (86.67\%) lncRNAs associated with breast cancer predicted by DBNLDA were also confirmed by lnc2cancer or LncRNADisease database. For the two unconfirmed predictions, we could find the evidence from recent publications. In the case of lung cancer, among the top 15 predicted lncRNAs, 10 were  confirmed by reference databases and remaining were reported in recent literature. In case of stomach cancer, DBNLDA could predict 12 associations reported either by reference databases or literature. The three new associations (HCP5, HCG4 and MIR99AHG, indicated by '*' in Table~\ref{Tab:cancer3}) which could be considered as new suggestions for further laboratory validations. The detailed comparison of LDA prediction by DBNLDA is available in SupplementaryTable-S2.xls.

\section{Conclusion} \label{conclusion}
Identifying disease-associated lncRNA is a necessary step in discerning the functional roles of lncRNAs in the disease mechanism. In this work, we developed DBNLDA, a deep belief network-based model for lncRNA-disease association prediction. This work integrates information of lncRNA, miRNA,  disease interactions, and functional similarities to construct heterogeneous networks. Then DBN based latent representations of lncRNAs and diseases are used to predict the lncRNA-disease association accurately. The cross-validation confirmed that DBNLDA has comparable performance in terms of AUC and significant improvement in terms of AUPR. Case studies on breast cancer, lung cancer and stomach cancer show the ability of DBNLDA to predict potential disease-associated lncRNAs. The model could be extended further with multi-modal data such as lncRNA drug-target interactions and lncRNA-epigenetic-disease interactions.  

\section*{Acknowledgement}

This research work is an outcome of the R\& D work under the Visvesvaraya PhD Scheme of Ministry of Electronics and Information Technology, Government of India \vspace*{-12pt}

\section*{References}


\begin{thebibliography}{44}
\expandafter\ifx\csname natexlab\endcsname\relax\def\natexlab#1{#1}\fi
\providecommand{\url}[1]{\texttt{#1}}
\providecommand{\href}[2]{#2}
\providecommand{\path}[1]{#1}
\providecommand{\DOIprefix}{doi:}
\providecommand{\ArXivprefix}{arXiv:}
\providecommand{\URLprefix}{URL: }
\providecommand{\Pubmedprefix}{pmid:}
\providecommand{\doi}[1]{\href{http://dx.doi.org/#1}{\path{#1}}}
\providecommand{\Pubmed}[1]{\href{pmid:#1}{\path{#1}}}
\providecommand{\bibinfo}[2]{#2}
\ifx\xfnm\relax \def\xfnm[#1]{\unskip,\space#1}\fi
\bibitem[{Goff and Rinn(2015)}]{goff2015linking}
\bibinfo{author}{L.~A. Goff}, \bibinfo{author}{J.~L. Rinn},
\newblock \bibinfo{title}{Linking rna biology to lncrnas},
\newblock \bibinfo{journal}{Genome research} \bibinfo{volume}{25}
  (\bibinfo{year}{2015}) \bibinfo{pages}{1456--1465}.
\bibitem[{Lee(2012)}]{lee2012epigenetic}
\bibinfo{author}{J.~T. Lee},
\newblock \bibinfo{title}{Epigenetic regulation by long noncoding rnas},
\newblock \bibinfo{journal}{Science} \bibinfo{volume}{338}
  (\bibinfo{year}{2012}) \bibinfo{pages}{1435--1439}.
\bibitem[{Huarte(2015)}]{huarte2015emerging}
\bibinfo{author}{M.~Huarte},
\newblock \bibinfo{title}{The emerging role of lncrnas in cancer},
\newblock \bibinfo{journal}{Nature medicine} \bibinfo{volume}{21}
  (\bibinfo{year}{2015}) \bibinfo{pages}{1253}.
\bibitem[{Luo and Chen(2016)}]{luo2016long}
\bibinfo{author}{Q.~Luo}, \bibinfo{author}{Y.~Chen},
\newblock \bibinfo{title}{Long noncoding rnas and alzheimer’s disease},
\newblock \bibinfo{journal}{Clinical interventions in aging}
  \bibinfo{volume}{11} (\bibinfo{year}{2016}) \bibinfo{pages}{867}.
\bibitem[{Wang et~al.(2019)Wang, Zheng, Zheng, Cao, Zhang, Sun, and
  Wu}]{wang2019construction}
\bibinfo{author}{G.~Wang}, \bibinfo{author}{X.~Zheng},
  \bibinfo{author}{Y.~Zheng}, \bibinfo{author}{R.~Cao},
  \bibinfo{author}{M.~Zhang}, \bibinfo{author}{Y.~Sun},
  \bibinfo{author}{J.~Wu},
\newblock \bibinfo{title}{Construction and analysis of the lncrna-mirna-mrna
  network based on competitive endogenous rna reveals functional genes in heart
  failure},
\newblock \bibinfo{journal}{Molecular medicine reports} \bibinfo{volume}{19}
  (\bibinfo{year}{2019}) \bibinfo{pages}{994--1003}.
\bibitem[{Bhartiya et~al.(2012)Bhartiya, Kapoor, Jalali, Sati, Kaushik,
  Sachidanandan, Sivasubbu, and Scaria}]{bhartiya2012conceptual}
\bibinfo{author}{D.~Bhartiya}, \bibinfo{author}{S.~Kapoor},
  \bibinfo{author}{S.~Jalali}, \bibinfo{author}{S.~Sati},
  \bibinfo{author}{K.~Kaushik}, \bibinfo{author}{C.~Sachidanandan},
  \bibinfo{author}{S.~Sivasubbu}, \bibinfo{author}{V.~Scaria},
\newblock \bibinfo{title}{Conceptual approaches for lncrna drug discovery and
  future strategies},
\newblock \bibinfo{journal}{Expert opinion on drug discovery}
  \bibinfo{volume}{7} (\bibinfo{year}{2012}) \bibinfo{pages}{503--513}.
\bibitem[{Sun et~al.(2014)Sun, Shi, Wang, Zhang, Liu, Wang, He, Hao, Liu, and
  Zhou}]{sun2014inferring}
\bibinfo{author}{J.~Sun}, \bibinfo{author}{H.~Shi}, \bibinfo{author}{Z.~Wang},
  \bibinfo{author}{C.~Zhang}, \bibinfo{author}{L.~Liu},
  \bibinfo{author}{L.~Wang}, \bibinfo{author}{W.~He}, \bibinfo{author}{D.~Hao},
  \bibinfo{author}{S.~Liu}, \bibinfo{author}{M.~Zhou},
\newblock \bibinfo{title}{Inferring novel lncrna--disease associations based on
  a random walk model of a lncrna functional similarity network},
\newblock \bibinfo{journal}{Molecular BioSystems} \bibinfo{volume}{10}
  (\bibinfo{year}{2014}) \bibinfo{pages}{2074--2081}.
\bibitem[{Chen et~al.(2016)Chen, You, Yan, and Gong}]{chen2016irwrlda}
\bibinfo{author}{X.~Chen}, \bibinfo{author}{Z.-H. You}, \bibinfo{author}{G.-Y.
  Yan}, \bibinfo{author}{D.-W. Gong},
\newblock \bibinfo{title}{Irwrlda: improved random walk with restart for
  lncrna-disease association prediction},
\newblock \bibinfo{journal}{Oncotarget} \bibinfo{volume}{7}
  (\bibinfo{year}{2016}) \bibinfo{pages}{57919}.
\bibitem[{Yu et~al.(2017)Yu, Fu, Lu, Ren, and Wang}]{yu2017brwlda}
\bibinfo{author}{G.~Yu}, \bibinfo{author}{G.~Fu}, \bibinfo{author}{C.~Lu},
  \bibinfo{author}{Y.~Ren}, \bibinfo{author}{J.~Wang},
\newblock \bibinfo{title}{Brwlda: bi-random walks for predicting lncrna-disease
  associations},
\newblock \bibinfo{journal}{Oncotarget} \bibinfo{volume}{8}
  (\bibinfo{year}{2017}) \bibinfo{pages}{60429}.
\bibitem[{Chen(2015)}]{chen2015katzlda}
\bibinfo{author}{X.~Chen},
\newblock \bibinfo{title}{Katzlda: Katz measure for the lncrna-disease
  association prediction},
\newblock \bibinfo{journal}{Scientific reports} \bibinfo{volume}{5}
  (\bibinfo{year}{2015}) \bibinfo{pages}{16840}.
\bibitem[{Paraskevopoulou and
  Hatzigeorgiou(2016)}]{paraskevopoulou2016analyzing}
\bibinfo{author}{M.~D. Paraskevopoulou}, \bibinfo{author}{A.~G. Hatzigeorgiou},
\newblock \bibinfo{title}{Analyzing mirna--lncrna interactions},
\newblock in: \bibinfo{booktitle}{long non-coding RNAs},
  \bibinfo{publisher}{Springer}, \bibinfo{year}{2016}, pp.
  \bibinfo{pages}{271--286}.
\bibitem[{Yao et~al.(2020)Yao, Zhan, Zhan, Kwoh, Li, and Wang}]{rflda}
\bibinfo{author}{D.~Yao}, \bibinfo{author}{X.~Zhan}, \bibinfo{author}{X.~Zhan},
  \bibinfo{author}{C.~K. Kwoh}, \bibinfo{author}{P.~Li},
  \bibinfo{author}{J.~Wang},
\newblock \bibinfo{title}{A random forest based computational model for
  predicting novel lncrna-disease associations},
\newblock \bibinfo{journal}{BMC bioinformatics} \bibinfo{volume}{21}
  (\bibinfo{year}{2020}) \bibinfo{pages}{1--18}.
\bibitem[{Lu et~al.(2018)Lu, Yang, Luo, Wu, Li, Pan, Li, and
  Wang}]{lu2018simclda}
\bibinfo{author}{C.~Lu}, \bibinfo{author}{M.~Yang}, \bibinfo{author}{F.~Luo},
  \bibinfo{author}{F.-X. Wu}, \bibinfo{author}{M.~Li},
  \bibinfo{author}{Y.~Pan}, \bibinfo{author}{Y.~Li}, \bibinfo{author}{J.~Wang},
\newblock \bibinfo{title}{Prediction of lncrna--disease associations based on
  inductive matrix completion},
\newblock \bibinfo{journal}{Bioinformatics} \bibinfo{volume}{34}
  (\bibinfo{year}{2018}) \bibinfo{pages}{3357--3364}.
\bibitem[{Fu et~al.(2018)Fu, Wang, Domeniconi, and Yu}]{fu2018mflda}
\bibinfo{author}{G.~Fu}, \bibinfo{author}{J.~Wang},
  \bibinfo{author}{C.~Domeniconi}, \bibinfo{author}{G.~Yu},
\newblock \bibinfo{title}{Matrix factorization-based data fusion for the
  prediction of lncrna--disease associations},
\newblock \bibinfo{journal}{Bioinformatics} \bibinfo{volume}{34}
  (\bibinfo{year}{2018}) \bibinfo{pages}{1529--1537}.
\bibitem[{Ding et~al.(2018)Ding, Wang, Sun, and Li}]{ding2018tpglda}
\bibinfo{author}{L.~Ding}, \bibinfo{author}{M.~Wang}, \bibinfo{author}{D.~Sun},
  \bibinfo{author}{A.~Li},
\newblock \bibinfo{title}{Tpglda: Novel prediction of associations between
  lncrnas and diseases via lncrna-disease-gene tripartite graph},
\newblock \bibinfo{journal}{Scientific reports} \bibinfo{volume}{8}
  (\bibinfo{year}{2018}) \bibinfo{pages}{1--11}.
\bibitem[{Wang et~al.(2017)Wang, Guo, Gao, Zhi, Zhang, Liu, Zhang, Yue, Guo,
  Ning et~al.}]{wang2017improved}
\bibinfo{author}{P.~Wang}, \bibinfo{author}{Q.~Guo}, \bibinfo{author}{Y.~Gao},
  \bibinfo{author}{H.~Zhi}, \bibinfo{author}{Y.~Zhang},
  \bibinfo{author}{Y.~Liu}, \bibinfo{author}{J.~Zhang},
  \bibinfo{author}{M.~Yue}, \bibinfo{author}{M.~Guo},
  \bibinfo{author}{S.~Ning}, et~al.,
\newblock \bibinfo{title}{Improved method for prioritization of disease
  associated lncrnas based on cerna theory and functional genomics data},
\newblock \bibinfo{journal}{Oncotarget} \bibinfo{volume}{8}
  (\bibinfo{year}{2017}) \bibinfo{pages}{4642}.
\bibitem[{Lan et~al.(2017)Lan, Li, Zhao, Liu, Wu, Pan, and Wang}]{lan2017ldap}
\bibinfo{author}{W.~Lan}, \bibinfo{author}{M.~Li}, \bibinfo{author}{K.~Zhao},
  \bibinfo{author}{J.~Liu}, \bibinfo{author}{F.-X. Wu},
  \bibinfo{author}{Y.~Pan}, \bibinfo{author}{J.~Wang},
\newblock \bibinfo{title}{Ldap: a web server for lncrna-disease association
  prediction},
\newblock \bibinfo{journal}{Bioinformatics} \bibinfo{volume}{33}
  (\bibinfo{year}{2017}) \bibinfo{pages}{458--460}.
\bibitem[{Chen et~al.(2015)Chen, Yan, Luo, Ji, Zhang, and Dai}]{lrlslda}
\bibinfo{author}{X.~Chen}, \bibinfo{author}{C.~C. Yan},
  \bibinfo{author}{C.~Luo}, \bibinfo{author}{W.~Ji},
  \bibinfo{author}{Y.~Zhang}, \bibinfo{author}{Q.~Dai},
\newblock \bibinfo{title}{Constructing lncrna functional similarity network
  based on lncrna-disease associations and disease semantic similarity},
\newblock \bibinfo{journal}{Scientific reports} \bibinfo{volume}{5}
  (\bibinfo{year}{2015}) \bibinfo{pages}{11338}.
\bibitem[{Xuan et~al.(2019{\natexlab{a}})Xuan, Cao, Zhang, Kong, and
  Zhang}]{xuan2019dual}
\bibinfo{author}{P.~Xuan}, \bibinfo{author}{Y.~Cao},
  \bibinfo{author}{T.~Zhang}, \bibinfo{author}{R.~Kong},
  \bibinfo{author}{Z.~Zhang},
\newblock \bibinfo{title}{Dual convolutional neural networks with attention
  mechanisms based method for predicting disease - related lncrna genes},
\newblock \bibinfo{journal}{Frontiers in genetics} \bibinfo{volume}{10}
  (\bibinfo{year}{2019}{\natexlab{a}}) \bibinfo{pages}{416}.
\bibitem[{Xuan et~al.(2019{\natexlab{b}})Xuan, Pan, Zhang, Liu, and
  Sun}]{gcnlda}
\bibinfo{author}{P.~Xuan}, \bibinfo{author}{S.~Pan},
  \bibinfo{author}{T.~Zhang}, \bibinfo{author}{Y.~Liu},
  \bibinfo{author}{H.~Sun},
\newblock \bibinfo{title}{Graph convolutional network and convolutional neural
  network based method for predicting lncrna-disease associations},
\newblock \bibinfo{journal}{Cells} \bibinfo{volume}{8}
  (\bibinfo{year}{2019}{\natexlab{b}}) \bibinfo{pages}{1012}.
\bibitem[{Kipf and Welling(2016)}]{kipf2016semi}
\bibinfo{author}{T.~N. Kipf}, \bibinfo{author}{M.~Welling},
\newblock \bibinfo{title}{Semi-supervised classification with graph
  convolutional networks},
\newblock \bibinfo{journal}{arXiv preprint arXiv:1609.02907}
  (\bibinfo{year}{2016}).
\bibitem[{Wu et~al.(2020)Wu, Lan, Chen, Dong, Liu, and Peng}]{wu2020inferring}
\bibinfo{author}{X.~Wu}, \bibinfo{author}{W.~Lan}, \bibinfo{author}{Q.~Chen},
  \bibinfo{author}{Y.~Dong}, \bibinfo{author}{J.~Liu},
  \bibinfo{author}{W.~Peng},
\newblock \bibinfo{title}{Inferring lncrna-disease associations based on graph
  autoencoder matrix completion},
\newblock \bibinfo{journal}{Computational Biology and Chemistry}
  (\bibinfo{year}{2020}) \bibinfo{pages}{107282}.
\bibitem[{Hinton(2009)}]{hinton2009deep}
\bibinfo{author}{G.~E. Hinton},
\newblock \bibinfo{title}{Deep belief networks},
\newblock \bibinfo{journal}{Scholarpedia} \bibinfo{volume}{4}
  (\bibinfo{year}{2009}) \bibinfo{pages}{5947}.
\bibitem[{Wang and Zeng(2013)}]{wang2013predicting}
\bibinfo{author}{Y.~Wang}, \bibinfo{author}{J.~Zeng},
\newblock \bibinfo{title}{Predicting drug-target interactions using restricted
  boltzmann machines},
\newblock \bibinfo{journal}{Bioinformatics} \bibinfo{volume}{29}
  (\bibinfo{year}{2013}) \bibinfo{pages}{i126--i134}.
\bibitem[{Chen et~al.(2015)Chen, Yan, Zhang, Li, Deng, Zhang, and
  Dai}]{chen2015rbmmmda}
\bibinfo{author}{X.~Chen}, \bibinfo{author}{C.~C. Yan},
  \bibinfo{author}{X.~Zhang}, \bibinfo{author}{Z.~Li},
  \bibinfo{author}{L.~Deng}, \bibinfo{author}{Y.~Zhang},
  \bibinfo{author}{Q.~Dai},
\newblock \bibinfo{title}{Rbmmmda: predicting multiple types of
  disease-microrna associations},
\newblock \bibinfo{journal}{Scientific reports} \bibinfo{volume}{5}
  (\bibinfo{year}{2015}) \bibinfo{pages}{13877}.
\bibitem[{Luo et~al.(2019)Luo, Li, Tian, and Wu}]{dgmdl}
\bibinfo{author}{P.~Luo}, \bibinfo{author}{Y.~Li}, \bibinfo{author}{L.-P.
  Tian}, \bibinfo{author}{F.-X. Wu},
\newblock \bibinfo{title}{Enhancing the prediction of disease--gene
  associations with multimodal deep learning},
\newblock \bibinfo{journal}{Bioinformatics} \bibinfo{volume}{35}
  (\bibinfo{year}{2019}) \bibinfo{pages}{3735--3742}.
\bibitem[{Liang et~al.(2014)Liang, Li, Chen, and Zeng}]{liang2014integrative}
\bibinfo{author}{M.~Liang}, \bibinfo{author}{Z.~Li}, \bibinfo{author}{T.~Chen},
  \bibinfo{author}{J.~Zeng},
\newblock \bibinfo{title}{Integrative data analysis of multi-platform cancer
  data with a multimodal deep learning approach},
\newblock \bibinfo{journal}{IEEE/ACM transactions on computational biology and
  bioinformatics} \bibinfo{volume}{12} (\bibinfo{year}{2014})
  \bibinfo{pages}{928--937}.
\bibitem[{Chen et~al.(2012)Chen, Wang, Wang, Qiu, Liu, Chen, Zhang, Yan, and
  Cui}]{chen2012lncrnadisease}
\bibinfo{author}{G.~Chen}, \bibinfo{author}{Z.~Wang},
  \bibinfo{author}{D.~Wang}, \bibinfo{author}{C.~Qiu},
  \bibinfo{author}{M.~Liu}, \bibinfo{author}{X.~Chen},
  \bibinfo{author}{Q.~Zhang}, \bibinfo{author}{G.~Yan},
  \bibinfo{author}{Q.~Cui},
\newblock \bibinfo{title}{Lncrnadisease: a database for long-non-coding
  rna-associated diseases},
\newblock \bibinfo{journal}{Nucleic acids research} \bibinfo{volume}{41}
  (\bibinfo{year}{2012}) \bibinfo{pages}{D983--D986}.
\bibitem[{Ning et~al.(2016)Ning, Zhang, Wang, Zhi, Wang, Liu, Gao, Guo, Yue,
  Wang et~al.}]{ning2016lnc2cancer}
\bibinfo{author}{S.~Ning}, \bibinfo{author}{J.~Zhang},
  \bibinfo{author}{P.~Wang}, \bibinfo{author}{H.~Zhi},
  \bibinfo{author}{J.~Wang}, \bibinfo{author}{Y.~Liu},
  \bibinfo{author}{Y.~Gao}, \bibinfo{author}{M.~Guo}, \bibinfo{author}{M.~Yue},
  \bibinfo{author}{L.~Wang}, et~al.,
\newblock \bibinfo{title}{Lnc2cancer: a manually curated database of
  experimentally supported lncrnas associated with various human cancers},
\newblock \bibinfo{journal}{Nucleic acids research} \bibinfo{volume}{44}
  (\bibinfo{year}{2016}) \bibinfo{pages}{D980--D985}.
\bibitem[{Fan et~al.(2016)Fan, Siklenka, Arora, Ribeiro, Kimmins, and
  Xia}]{fan2016mirnet}
\bibinfo{author}{Y.~Fan}, \bibinfo{author}{K.~Siklenka}, \bibinfo{author}{S.~K.
  Arora}, \bibinfo{author}{P.~Ribeiro}, \bibinfo{author}{S.~Kimmins},
  \bibinfo{author}{J.~Xia},
\newblock \bibinfo{title}{mirnet-dissecting mirna-target interactions and
  functional associations through network-based visual analysis},
\newblock \bibinfo{journal}{Nucleic acids research} \bibinfo{volume}{44}
  (\bibinfo{year}{2016}) \bibinfo{pages}{W135--W141}.
\bibitem[{Li et~al.(2014)Li, Liu, Zhou, Qu, and Yang}]{li2014starbase}
\bibinfo{author}{J.-H. Li}, \bibinfo{author}{S.~Liu},
  \bibinfo{author}{H.~Zhou}, \bibinfo{author}{L.-H. Qu}, \bibinfo{author}{J.-H.
  Yang},
\newblock \bibinfo{title}{starbase v2. 0: decoding mirna-cerna, mirna-ncrna and
  protein--rna interaction networks from large-scale clip-seq data},
\newblock \bibinfo{journal}{Nucleic acids research} \bibinfo{volume}{42}
  (\bibinfo{year}{2014}) \bibinfo{pages}{D92--D97}.
\bibitem[{Chen et~al.(2015)Chen, Yan, Luo, Ji, Zhang, and
  Dai}]{chen2015constructing}
\bibinfo{author}{X.~Chen}, \bibinfo{author}{C.~C. Yan},
  \bibinfo{author}{C.~Luo}, \bibinfo{author}{W.~Ji},
  \bibinfo{author}{Y.~Zhang}, \bibinfo{author}{Q.~Dai},
\newblock \bibinfo{title}{Constructing lncrna functional similarity network
  based on lncrna-disease associations and disease semantic similarity},
\newblock \bibinfo{journal}{Scientific reports} \bibinfo{volume}{5}
  (\bibinfo{year}{2015}) \bibinfo{pages}{11338}.
\bibitem[{Grover and Leskovec(2016)}]{grover2016node2vec}
\bibinfo{author}{A.~Grover}, \bibinfo{author}{J.~Leskovec},
\newblock \bibinfo{title}{node2vec: Scalable feature learning for networks},
\newblock in: \bibinfo{booktitle}{Proceedings of the 22nd ACM SIGKDD
  international conference on Knowledge discovery and data mining},
  \bibinfo{year}{2016}, pp. \bibinfo{pages}{855--864}.
\bibitem[{Ping et~al.(2018)Ping, Wang, Kuang, Ye, Iqbal, and
  Pei}]{ping2018novel}
\bibinfo{author}{P.~Ping}, \bibinfo{author}{L.~Wang},
  \bibinfo{author}{L.~Kuang}, \bibinfo{author}{S.~Ye},
  \bibinfo{author}{M.~F.~B. Iqbal}, \bibinfo{author}{T.~Pei},
\newblock \bibinfo{title}{A novel method for lncrna-disease association
  prediction based on an lncrna-disease association network},
\newblock \bibinfo{journal}{IEEE/ACM transactions on computational biology and
  bioinformatics} \bibinfo{volume}{16} (\bibinfo{year}{2018})
  \bibinfo{pages}{688--693}.
\bibitem[{Wu et~al.(2019)Wu, Chen, Ye, Wang, Zhang, Sheng, Meng, and
  Chen}]{wu2019downregulation}
\bibinfo{author}{J.~Wu}, \bibinfo{author}{H.~Chen}, \bibinfo{author}{M.~Ye},
  \bibinfo{author}{B.~Wang}, \bibinfo{author}{Y.~Zhang},
  \bibinfo{author}{J.~Sheng}, \bibinfo{author}{T.~Meng},
  \bibinfo{author}{H.~Chen},
\newblock \bibinfo{title}{Downregulation of long noncoding rna hcp5 contributes
  to cisplatin resistance in human triple-negative breast cancer via regulation
  of pten expression},
\newblock \bibinfo{journal}{Biomedicine \& Pharmacotherapy}
  \bibinfo{volume}{115} (\bibinfo{year}{2019}) \bibinfo{pages}{108869}.
\bibitem[{Wang et~al.(2018)Wang, Ke, Zhang, Ma, Ao, Zou, Yang, Zhu, Nie, Wu
  et~al.}]{wang2018lncrna}
\bibinfo{author}{S.~Wang}, \bibinfo{author}{H.~Ke}, \bibinfo{author}{H.~Zhang},
  \bibinfo{author}{Y.~Ma}, \bibinfo{author}{L.~Ao}, \bibinfo{author}{L.~Zou},
  \bibinfo{author}{Q.~Yang}, \bibinfo{author}{H.~Zhu},
  \bibinfo{author}{J.~Nie}, \bibinfo{author}{C.~Wu}, et~al.,
\newblock \bibinfo{title}{Lncrna mir100hg promotes cell proliferation in
  triple-negative breast cancer through triplex formation with p27 loci},
\newblock \bibinfo{journal}{Cell death \& disease} \bibinfo{volume}{9}
  (\bibinfo{year}{2018}) \bibinfo{pages}{1--11}.
\bibitem[{Guo et~al.(2019)Guo, Zhang, Zhang, Wu, He, Li, and
  Wang}]{guo2019long}
\bibinfo{author}{S.~Guo}, \bibinfo{author}{L.~Zhang},
  \bibinfo{author}{Y.~Zhang}, \bibinfo{author}{Z.~Wu}, \bibinfo{author}{D.~He},
  \bibinfo{author}{X.~Li}, \bibinfo{author}{Z.~Wang},
\newblock \bibinfo{title}{Long non-coding rna tug1 enhances chemosensitivity in
  non-small cell lung cancer by impairing microrna-221-dependent pten
  inhibition},
\newblock \bibinfo{journal}{Aging (Albany NY)} \bibinfo{volume}{11}
  (\bibinfo{year}{2019}) \bibinfo{pages}{7553}.
\bibitem[{Wang et~al.(2017)Wang, Shen, Zhang, Yang, Cui, Sun, Wang, Fan, and
  Xu}]{wang2017long}
\bibinfo{author}{H.~Wang}, \bibinfo{author}{Q.~Shen},
  \bibinfo{author}{X.~Zhang}, \bibinfo{author}{C.~Yang},
  \bibinfo{author}{S.~Cui}, \bibinfo{author}{Y.~Sun},
  \bibinfo{author}{L.~Wang}, \bibinfo{author}{X.~Fan}, \bibinfo{author}{S.~Xu},
\newblock \bibinfo{title}{The long non-coding rna xist controls non-small cell
  lung cancer proliferation and invasion by modulating mir-186-5p},
\newblock \bibinfo{journal}{Cellular Physiology and Biochemistry}
  \bibinfo{volume}{41} (\bibinfo{year}{2017}) \bibinfo{pages}{2221--2229}.
\bibitem[{Liu et~al.(2018)Liu, Zhou, Zhang, Wang, He, Chen, Huang, Li, and
  Li}]{liu2018lncrna}
\bibinfo{author}{L.~Liu}, \bibinfo{author}{X.~Zhou},
  \bibinfo{author}{J.~Zhang}, \bibinfo{author}{G.~Wang},
  \bibinfo{author}{J.~He}, \bibinfo{author}{Y.~Chen},
  \bibinfo{author}{C.~Huang}, \bibinfo{author}{L.~Li}, \bibinfo{author}{S.~Li},
\newblock \bibinfo{title}{Lncrna hulc promotes non-small cell lung cancer cell
  proliferation and inhibits the apoptosis by up-regulating sphingosine kinase
  1 (sphk1) and its downstream pi3k/akt pathway},
\newblock \bibinfo{journal}{Eur. Rev. Med. Pharmacol. Sci} \bibinfo{volume}{22}
  (\bibinfo{year}{2018}) \bibinfo{pages}{8722--8730}.
\bibitem[{Song et~al.(2020)Song, Wang, and Zong}]{song2020lncrna}
\bibinfo{author}{J.~Song}, \bibinfo{author}{Q.~Wang},
  \bibinfo{author}{L.~Zong},
\newblock \bibinfo{title}{Lncrna mir155hg contributes to smoke-related chronic
  obstructive pulmonary disease by targeting mir-128-5p/brd4 axis},
\newblock \bibinfo{journal}{Bioscience Reports} \bibinfo{volume}{40}
  (\bibinfo{year}{2020}).
\bibitem[{Du et~al.(2018)Du, Hao, and Liu}]{du2018low}
\bibinfo{author}{Y.~Du}, \bibinfo{author}{X.~Hao}, \bibinfo{author}{X.~Liu},
\newblock \bibinfo{title}{Low expression of long noncoding rna cdkn2b-as1 in
  patients with idiopathic pulmonary fibrosis predicts lung cancer by
  regulating the p53-signaling pathway},
\newblock \bibinfo{journal}{Oncology letters} \bibinfo{volume}{15}
  (\bibinfo{year}{2018}) \bibinfo{pages}{4912--4918}.
\bibitem[{Bahari et~al.(2015)Bahari, Emadi-Baygi, Nikpour
  et~al.}]{bahari2015mir}
\bibinfo{author}{F.~Bahari}, \bibinfo{author}{M.~Emadi-Baygi},
  \bibinfo{author}{P.~Nikpour}, et~al.,
\newblock \bibinfo{title}{mir-17-92 host gene, uderexpressed in gastric cancer
  and its expression was negatively correlated with the metastasis},
\newblock \bibinfo{journal}{Indian journal of cancer} \bibinfo{volume}{52}
  (\bibinfo{year}{2015}) \bibinfo{pages}{22}.
\bibitem[{Ren et~al.(2018)Ren, Yang, Yang, Zhang, Zhao, Wei, Zhang, and
  Zhang}]{ren2018upregulation}
\bibinfo{author}{H.~Ren}, \bibinfo{author}{X.~Yang}, \bibinfo{author}{Y.~Yang},
  \bibinfo{author}{X.~Zhang}, \bibinfo{author}{R.~Zhao},
  \bibinfo{author}{R.~Wei}, \bibinfo{author}{X.~Zhang},
  \bibinfo{author}{Y.~Zhang},
\newblock \bibinfo{title}{Upregulation of lncrna bcyrn1 promotes tumor
  progression and enhances epcam expression in gastric carcinoma},
\newblock \bibinfo{journal}{Oncotarget} \bibinfo{volume}{9}
  (\bibinfo{year}{2018}) \bibinfo{pages}{4851}.
\bibitem[{Piao et~al.(2019)Piao, Guo, Wang, and Zhang}]{piao2019long}
\bibinfo{author}{H.-Y. Piao}, \bibinfo{author}{S.~Guo},
  \bibinfo{author}{Y.~Wang}, \bibinfo{author}{J.~Zhang},
\newblock \bibinfo{title}{Long noncoding rna nalt1-induced gastric cancer
  invasion and metastasis via notch signaling pathway},
\newblock \bibinfo{journal}{World Journal of Gastroenterology}
  \bibinfo{volume}{25} (\bibinfo{year}{2019}) \bibinfo{pages}{6508}.

\end{thebibliography}

\end{document}